\documentclass[%
 reprint,
 amsmath,amssymb,
 aps,
]{revtex4-1}

\usepackage{graphicx}
\usepackage{dcolumn}
\usepackage{bm}
\usepackage[mathlines]{lineno}

\begin{document}

\preprint{APS/123-QED}

\title{Coupling dark-baryonic matter density profile for vacuum decay scenarios}
\thanks{A footnote to the article title}%

\author{H. T. C. M. Souza}
\affiliation{Universidade Federal Rural do Semi-\'Arido - DECEN, Brazil} 
 \affiliation{Universidade Federal do Rio Grande do Norte - DFTE, Brazil}
  \email{hidalyn.souza@ufersa.edu.br}
\author{N. Pires}%
 \email{npires@dfte.ufrn.br}
 \affiliation{Universidade Federal do Rio Grande do Norte - DFTE, Brazil}%
\author{F. E. M. Costa}\email{ernandesmatos@ufersa.edu.br}
\affiliation{Universidade Federal Rural do Semi-\'Arido - DECEN, Brazil}
 \author{F. C. Carvalho}%
 \email{fabiocabral@uern.br}
\author{E. P. Bento}%
 \email{eliangela@dfte.ufrn.br}
 \affiliation{Universidade do Estado do Rio Grande do Norte - DF, Brazil}%

\date{\today}

\begin{abstract}

The cosmological consequences of an interacting model in which vacuum decay law is deducted from the effect that vacuum decay has on the dark matter evolution are investigated. Here, the baryonic matter is also considered as a fluid gravitationally coupled with dark matter. It is made a careful analysis to constrain this model with the observational data of growth rate of cosmic structures. The theoretical growth rate is followed since the primordial recombination and the main physical processes on the baryonic component are considered. As a complementary constraint, this model is compared with the observed CMB-BAO ratio as well with the gas mass fraction of cluster of galaxies. We found the best fit values for dark matter $\Omega_{d0} = 0.269  ^{+0.023}_{-0.023}$ and for the decay parameter $\epsilon = 0.02 ^{+0.04}_{-0.05}$.
\begin{description}
\item[Keywords]
Spherical Top-Hat collapse, linear growth rate, cooling function, vacuum decay model.
\end{description}
\end{abstract}

\pacs{Valid PACS appear here}
\maketitle


\section{\label{sec:intro}Introduction}

Several astronomical observations (Supernova Ia distance, age of the universe estimates, measurements of the cosmic microwave background radiation anisotropies etc.) have convergently indicated that the universe is undergoing by a recent accelerated expansion. In the context of the General Relativity, this has been attributed to an exotic component of energy with negative pressure, which is the so called dark energy (see a pedagogical review in \cite{Frieman2009}).

The nature and origin of this dark component is still a completely open question. Among several candidates for dark energy, the oldest and most natural is the cosmological constant $\Lambda$ \cite{Weinberg1989_61}, interpreted as the vacuum energy, i.e., $ \rho_\Lambda = \Lambda/8\pi G$.
Moreover, here we face with the cosmological constant problem which is the conflict between the upper cosmological limit today ($ \rho_\Lambda\sim 10^{-47} GeV^4$), and the theoretical expectation within the framework of the quantum field theory ($\sim 10^{71} GeV^4$) for the vacuum energy value. In this regard, a phenomenological attempt to alleviate such  problem is allowing $\Lambda$ to vary with time. Even
before the discovery of the accelerating expansion, many works had been already dedicated in the time-evolving vacuum models. These models were motivated in the attempt to explain the cosmological constant problem as well as the age of the universe \cite{Ozer1986_171}. With the discovery of the accelerated expansion the interest in them increase, because they can explain the accelerated expansion of the universe in an efficient way, and also provide an interesting attempt to evade the coincidence and cosmological constant problems of the standard $\Lambda$-cosmology (\cite{Shapiro2000_475}, and also for a more complete history see \cite{Lima2013_a}).

An interesting and more realistic approach based on the modified matter expansion rate was proposed by Wang $\&$ Meng \cite{Wang2005_22}. Afterwards, in this model, Alcaniz $\&$ Lima \cite{Alcaniz2005_72} included the baryonic component and showed that its presence changes the dynamic of the evolution altering the transition redshift for values compatible with current estimates based on type Ia supernova.

The aim of the present paper is to investigate the cosmology of the vacuum decaying into cold dark matter (CDM), proposed in \cite{Wang2005_22}, considering the growth rate of matter (dark and baryonic) since the recombination era taking into account the main physical mechanisms in the baryonic component, through the exact calculations of the hydrodynamical equations. The physical processes that we take into account are photon drag, photon cooling, recombination, photoionization, collisional ionization, and hydrogen molecule production, destruction and cooling - hereafter called simply `physical mechanisms'. Baryonic and dark matter are considered as two fluids coupled only by gravity.

The outline of our paper is as follows. In section~\ref{VDM} is briefly described the vacuum decay model. Our model of dark and baryonic contrast and the growth rate in framework proposed by \cite{Wang2005_22} are presented in section~\ref{CC}, as well the constraint with the observational data of growth rate of cosmological structures. As an additional constraint for the vacuum decay model are done two complementary observation tests: ``CMB-BAO'' and ``Gas mass fraction of galaxy clusters'', presented in section~\ref{Constraints}. The conclusions are presented in section~\ref{Conc}, and in appendix~\ref{Ap_A} is showed the calculations of the temporal evolution of the baryonic matter.

\section{\label{VDM} Vacuum decay model}

Let us first consider a homogeneous and isotropic universe described by the Friedmann-Lemaitre-Robertson-Walker element line and a coupling between $\Lambda$ and CDM particles. In this case, we have that
\begin{equation}\label{ec}
\dot{\rho}_{d} + 3\frac{\dot{a}}{a}\rho_{d} = - \dot{\rho}_\Lambda\;,
\end{equation}
where $\rho_{d}$ and $\rho_\Lambda$ are the energy densities of CDM and $\Lambda$, respectively, $a$ is the cosmic scale factor and the dot sign denotes derivative with respect to the time.

In the traditional approach, the description of the cosmological scenario with a dynamical $\Lambda$ term there is the necessity of specify a phenomenological time dependent law for the vacuum. However, Ref. \cite{Wang2005_22} (see also \cite{Alcaniz2005_72}) found a decay law is deducted from the effect that vacuum has on the dark matter evolution.

The qualitative argument used in Ref. \cite{Wang2005_22} is the following: if vacuum is decaying into CDM particles, the energy density of this latter component will dilute more slowly compared to its standard evolution $\rho_{_d} \propto a^{-3}$. Thus, the deviation from the standard dilution is characterized by a positive constant, such that
\begin{equation}\label{rod}
\rho_{_d}= \rho_{_{d0}} a^{-3+\epsilon} \ ,
\end{equation}
where $\rho_{_{d0}}$ is the current value of $\rho_d$ (for the remainder of this and the next sections, all quantities with subscript zero will imply that we are considering the current value).

Now, by substituting the ansatz \eqref{rod} in Eq. \eqref{ec} yields the vacuum energy density
\begin{equation}\label{rovac}
\rho_{_v}=\tilde{\rho}_{_{v0}}+\frac{\epsilon
       \rho_{_{d0}}}{3-\epsilon}a^{-3+\epsilon} \ ,
\end{equation}
where $\tilde{ \rho}_{_{v0}}$ is an integration constant which represents the `ground state value of the vacuum', i.e., the standard vacuum energy density. Notice that when $\epsilon=0$ the standard $\Lambda CDM$ model is recovered.

Neglecting the radiation contribution and considering that the baryonic content is separately conserved, the expansion rate of the Universe can be written as
\begin{equation}\label{H}
H=H_{_0}\left[\Omega_{_{b0}}a^{-3}+\frac{3 \Omega_{_{d0}}}{3-\epsilon}a^{-3+\epsilon}+\tilde{\Omega}_{_{v0}}\right]^{1/2} \ ,
\end{equation}
where $\Omega_{b0} = \rho_{b0}/\rho_{c0}$ and $\Omega_{d0} = \rho_{d0}/\rho_{c0}$ represent the density parameters of baryons and dark matter, while that $\tilde{\Omega}_{_{v0}}=\tilde{\rho}_{_{v0}}/\rho_{_{c0}}$ the density parameter associated to constant $\tilde{\rho}_{_{v0}}$. From normalization condiction, one has
\begin{equation}\label{normalization}
\tilde{\Omega}_{_{v0}}= 1 - \Omega_{_{b0}} - \frac{3 \Omega_{_{d0}}}{3-\epsilon}\;\;.
\end{equation}
The term $\Omega_{b0}$ in Eq. \eqref{H}, separately conserved, was included by Ref. \cite{Alcaniz2005_72}, which showed that baryonic contribution cannot be neglected because it reconciles the decaying vacuum scenario with the observations. Besides, in the present work we are also taking into account the physical mechanisms in this component. 

As pointed in reference \cite{Alcaniz2005_72}, an important restriction on $\epsilon$ value given by second law of Thermodynamics implies that $\epsilon \geq 0$ and also $\epsilon \leq 1$ (or even should expect $\epsilon \ll 1$ due to no observational report anomalous on the CDM expansion rate yet), which avoid the accelerated phase of the universe has begun in the matter-dominated era \cite{Wang2005_22}. The analysis of the growth of perturbations can also provide one important restriction on the $\Lambda (t) CDM$ models. The presence of a dynamical vacuum energy provides a critical value to the decay parameter from which fluctuations do not collapse ($\epsilon>0.4$), no matter its size in the recombination. These  aspects will be considerably important because they provide an initial constraint when we will do the comparison with the observational data.

\section{\label{CC} Cosmological constraint: Growth rate of cosmic structure}
As we are focusing the importance of the attractive matter components, becomes interesting to compare this decaying model with the data set related with matter. In order to do this we compare our model with the observational growth factor (see table~\ref{tb02}).

To constrain the parametric space $\Omega_{_{d0}}-\epsilon$, we use the $\chi^2$ minimization

\begin{equation}\label{chi2gf}
\chi_{_{gf}}^2=\displaystyle\sum_{i=1}^{10}\left(\frac{f(z_{_i})_{_{obs}}-f(z_{_i},\theta_{_i})_{_{th}}}{\sigma_{_{gfi}}}\right)^2 \;\;,
\end{equation}
\\
where $f(z_{_i})_{_{th}}$ is the theoretical value of the growth factor given by equation \eqref{LGF} (bellow), $f(z_{_i})_{_{obs}}$ is the observational data shown in table~\ref{tb02} and $\sigma_{_{gfi}}$ is the error of the $f(z_{_i})_{_{obs}}$.
\begin{table}[!hptb]
\centering
\begin{tabular}{|l| c| c|} \hline
$~~~z$ &$f_{obs}$ &Ref. \\
\hline
0.02    &$0.48~\pm 0.09$     &\cite{Davis2011_413}\\
0.07    &$0.56~\pm 0.11$    &\cite{Beutler2012_423}\\
0.15    &$0.51~\pm 0.11$    &\cite{Hawkins2003_346}\\
0.22    &$0.60~\pm 0.10$    &\cite{Blake2011_415}\\
0.34    &$0.64~\pm 0.09$    &\cite{Cabre2009_393}\\
0.35    &$0.70~\pm 0.08$    &\cite{Tegmark2006_74}\\
0.41    &$0.70~\pm 0.07$    &\cite{Blake2011_415} \\
0.42    &$0.73~\pm 0.09$    &\cite{Blake2010_406} \\
0.59    &$0.75~\pm 0.09$    &\cite{Blake2010_406} \\
0.60    &$0.73~\pm 0.07$    &\cite{Blake2011_415}\\
0.78    &$0.70~\pm 0.08$    &\cite{Blake2011_415}\\
 \hline
 \end{tabular}
 \caption{\label{tb02} Observational values of the linear growth rate. In the first column is the sample redshift, the second one shows the growth rate value in the redshift and the corresponding central error bar (standard deviation) and, finally, in the third column the references related to each measurement.}
\end{table}

In the following, we present our theoretical model for the growth rate of the structures.

There is a tight connection between cold collisionless dark matter and baryonic matter. Furthermore, there are several physical mechanisms occurring in the baryonic matter which lack for dark matter during its evolution. The density contrast in these components were modelled as a cloud that contains a certain dark matter mass $M_d$ of radius $r_d$, and we follow only the baryonic matter that is inside this cloud. As described in appendix \ref{Ap_A}, the physical processes considered here are the photon cooling (heating), collisional ionization, photonionization, Lyman-$\alpha$ cooling, and the cooling due to $H_2$ formation. In order to evolve the matter fluctuation (or cloud), it was considered as having a `top-hat' profile as proposed in reference \cite{Opher1991_379}.

The baryonic matter and dark matter are considered as two fluids coupled only by gravity. As their evolution is non-relativistic, Newtonian hydrodynamic equations is used to describe them. The contrast densities for baryonic, $\delta_b$, and dark matter, $\delta_d$, are defined as usually,
\begin{equation}\label{contraste}
\delta_b=\frac{\delta{\rho_{_b}}}{\rho_{_b}} \qquad \mbox{and} \qquad \delta_d=\frac{\delta{\rho_{_d}}}{\rho_{_d}} .
\end{equation}

Considering the top-hat model, the  perturbation has a square density profile and a linear dependence for the velocity (which is consistent with the density profile)
\begin{subequations}\label{ps}
\begin{align}
\label{ps:1}
\rho_{ci}&=\rho_i+\rho_{1i}(t)=\rho_i[1+\delta_i(t)]
\\
\label{ps:2}
\textbf{v}_{ci}&=H\textbf{r}+v_{1i}(t)\displaystyle{\frac{\textbf{r}}{r_i}} \ ,
\end{align}
\end{subequations}
where the subscript $i$ means baryonic or dark matter, $r_i$ the radius of the clouds components, $\rho_{1i}(t)$ is the perturbation, $v_{1i}(t)$ is the peculiar velocity of the fluctuation, $\delta_{i}(t)$ is the density contrast of the perturbation, and $\textbf{r}$ is the radial coordinate centered in the cloud.

The collapse of a cloud with two components coupled initially begins with a cloud of dark and baryonic matter. As the collapse continues, the dark cloud collapses faster than the baryonic one, which is delayed by the physical mechanisms acting on it. Thus, part of the initial baryonic matter is left behind and we follow only that goes with the dark matter, i.e., $M_b(t)$. The surrounding medium forms a halo that could eventually interact with the cloud that is collapsing. As both baryonic and dark matter are into the same volume over time, the amount of baryonic matter $M_{_d}$ related to the mass of dark matter $M_{_d}$ shall be given by
\begin{equation}
M_{_b}(a)=M_{_d}\frac{\Omega_{_{b0}}}{\Omega_{_{d0}}}\left(\frac{1+\delta_{_b}}{1+\delta_{_d}}\right)a^{-\epsilon} \ .
\label{MB}
\end{equation}

The hydrodynamics equations are (see \cite{Opher1991_379} and \cite{Opher1997_285}) the continuity equation,
\begin{equation}\label{continuity}
\frac{\partial \rho_{_{ci}}}{\partial t} +  \nabla  \cdot \left(\rho_{_{ci}} \textbf{v}_{_{ci}}\right) = 0  \  ,
\end{equation}
the motion equation,
\begin{equation}\label{motion}
\frac{d\; \textbf{v}_{ci}}{dt}+(\textbf{v}_{ci}\cdot {\nabla})\textbf{v}_{ci}=-{\nabla}\phi-\frac{1}{\rho_{ci}}{\nabla} P-\frac{4}{3} \frac{\sigma_{_T}bT_{_\gamma}^4 x_{e}}{m_{p}c} (\textbf{v}_{ci}-H\textbf{r})\;\;,
\end{equation}
where $P$ is the pressure, $\sigma$ is the Thomson cross section, $b=4/c$ is the Stefan-Boltzmann's constant, $x_e$ is the degree of ionization, and the last term is the photon drag due to the background radiation; and the field equation itself is
\begin{equation}\label{field}
{\nabla}^2 \phi = 4 \pi G \rho_{ci} \ .
\end{equation}
Note that for the dark matter component the two last terms in equation \eqref{motion} do not exist (dark particles are collisionless and do not interact with photons).

Substituting the perturbations \eqref{contraste} in the subsequent equations \eqref{ps} - \eqref{field} we obtain the density contrast of dark matter at any time,
\begin{subequations}
\begin{equation}\label{DeltaPontodark}
\dot{\delta}_{d}+3(1+\delta_{d}) \frac{v_{1d}}{{r}_{d}}=0 \ ,
\end{equation}
and the velocity of the dark cloud
\begin{equation}
\label{EqVelDark}
\dot{v}_{_{1d}}+ Hv_{_{1d}} + \frac{4\pi G\rho_{_{0c}} {r}_{_d}}{3}
a^{-3}\left[\Omega_{_{b0}}+\Omega_{_{d0}}a^{\epsilon}\right](\delta_{_b}+\delta_{_d})=0 \, ;
\end{equation}
\end{subequations}
as well the evolution of the density contrast of the baryonic matter
\begin{subequations}
\begin{equation}\label{DeltaPontobary}
\dot{\delta}_{b}+3(1+\delta_{b}) \frac{v_{1b}}{{r}_{b}}=0 \ ,
\end{equation}
and the velocity of baryonic cloud inside the dark cloud (which follows the dark cloud),
\begin{eqnarray}\label{EqVelBari} \nonumber
\dot{v}_{_{1b}}+v_{_{1b}}&& \left(H+\frac{4}{3}\frac{\sigma_{_T} b T_{_{\gamma}}^4 x_{_e}}{m_{_{p}}\;c}\right)+\frac{4}{3} \pi G {r}_{_b}\rho_{_{c0}}a^{-3}\left[\Omega_{_{b0}}+\Omega_{_{d0}}a^{\epsilon}\right]\cdot\\
&&\cdot(\delta_{_b}+\delta_{_d}) -\frac{N_{_A}k_{_B}}{{r}_{_d}}T_{_m}(1+x_{_e})=0 .
\end{eqnarray}
\end{subequations}
Here, $T_m$ is the temperature of the baryonic matter, which has its dependence over time calculated in appendix \ref{Ap_A}, $m_p$ is the mass of proton and, just remembering, $\epsilon$ is the parameter of the vacuum decay.

According to equation \eqref{ps:2}, in the linear approximation the velocity of the fluctuations can be resolved into a term which tells us about the expansion of the universe, $H {\bf r}$, and also another representing the peculiar velocity of the cloud, as suggested by Peebles \cite{Peebles1980book}, $\textbf{u}(t)=v_{1i}(t)\textbf{r}/r_i=2/3[f(z) {\bf g}(\textbf{r})/(H(z)\Omega_{_m}(z))]$. In this expression $\textbf{g}(\textbf{r})$ is the specific gravity of the cloud and $f(z)$ is the linear growth rate, which is related with the growth mode of the perturbation of the baryonic and dark matter components as
\begin{equation}
f  = \frac{a}{\delta}\frac{d \delta}{da}=\frac{\dot{\delta_{b}}+\dot{\delta_{d}}}{\delta_{b}+\delta_{d}}\;\;.
\label{LGF}
\end{equation}
Originally, in 1980 Peebles found $f(z=0)=\Omega_{_{m0}}^{0.6}$ \cite{Peebles1980book}, a quite useful approximation of the linear growth rate obtained from the Friedmann-Lema\^itre model, valid for cases where the cosmological constant or curvature of space can be neglected \cite{Peebles}. Afterwards, more accurate approximations for this same model were found, as for instance, $f(z=0)=\Omega_{_{m0}}^{4/7}$ by Lightman $\&$ Schechter \cite{Lightman1990_74}. Later, other analysis were performed taking into account the cosmological constant and the variation of the growth rate with the redshift (for a brief review see for example  \cite{Opher1997_285}). For a flat universe with cosmological constant, a very useful utilized approximation was found by Lahav et al. \cite{Lahav1991_251}
\begin{equation}
f(z)=\Omega{'}_{_m}^{0.6}+\frac{1}{70}\left[1-\frac{\Omega'_{_m}}{2}(1+\Omega'_{_m})\right]\;\; .
\label{LahavEquation}
\end{equation}
where
\begin{equation}
\Omega'_{_m}(z) \equiv \frac{\Omega_{_{b0}}a^{-3}+\Omega_{_{d0}}a^{-3+\epsilon}}
{\Omega_{_{b0}}a^{-3}+\Omega_{_{d0}}a^{-3+\epsilon}+\Omega_{_{v0}}}\;\;, \label{Omega_m_bar}
\end{equation}
is the normalized matter density parameter. A similar approach was first made by Wang $\&$ Steinhardt \cite{Wang1998_508}.

\begin{figure}[hptb]
\vspace{0.5truecm}
\centering 
\hspace{0.25 truecm}\includegraphics[width=.45\textwidth,trim=90 90 90 90]{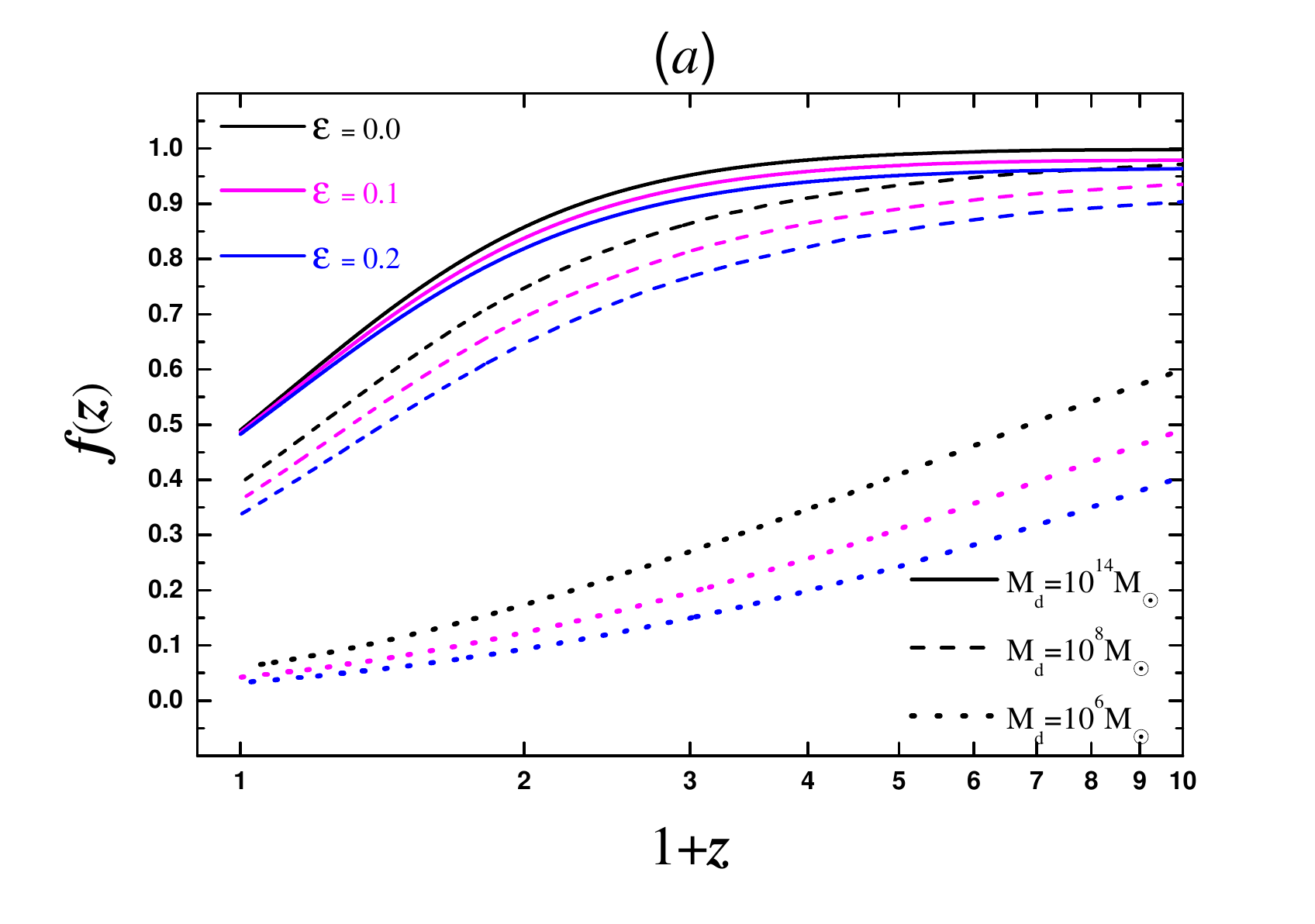}
\hfill \vspace{2.truecm}
\includegraphics[width=.45\textwidth,trim=90 90 90 90]{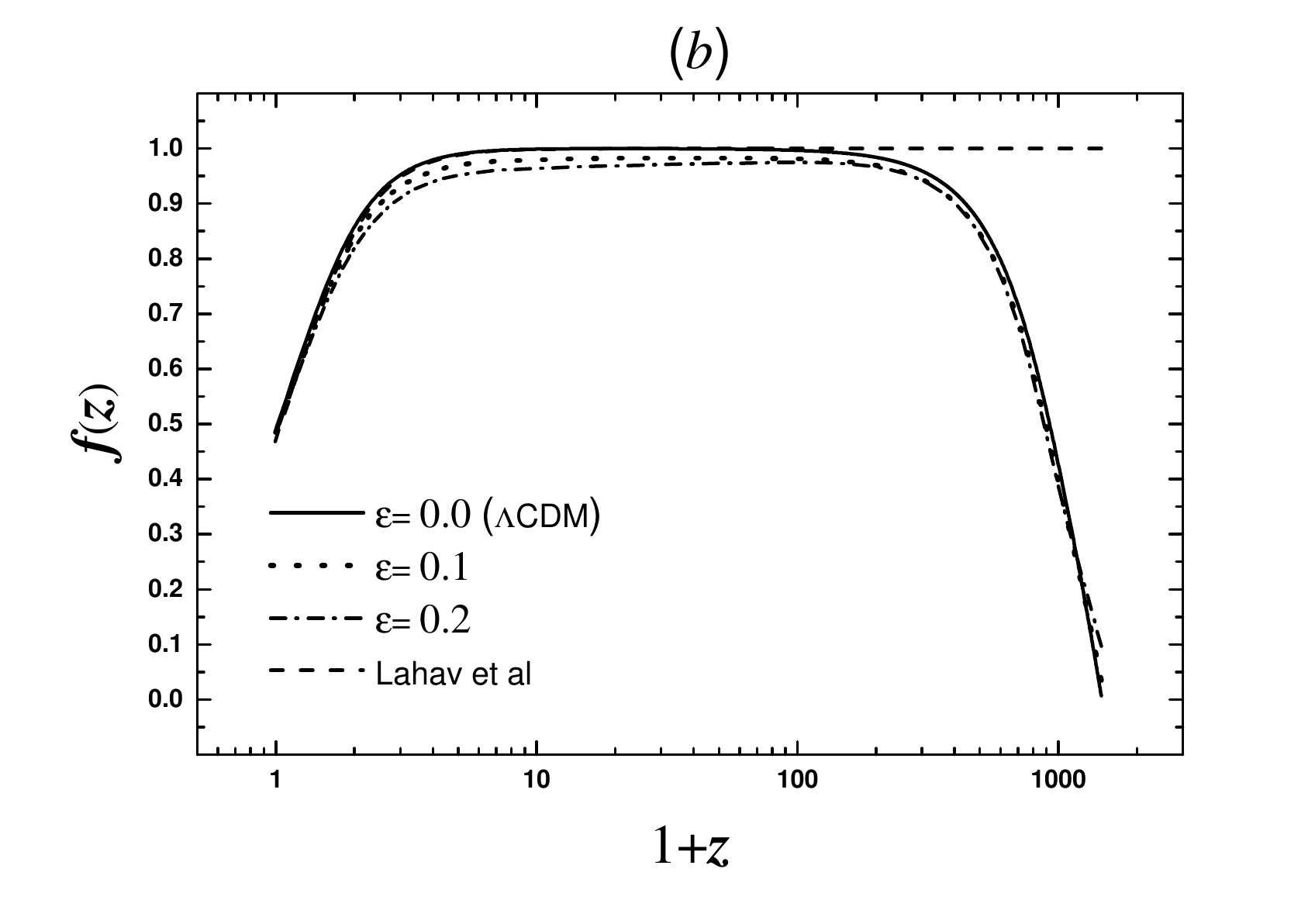}
\vspace{.6truecm}
\caption{\label{FigGrowth1} (a) Numerical evolution of the linear growth rate from equation~\eqref{LGF} for $\epsilon=0$ ($\equiv \Lambda$CDM), $\epsilon=0.1$, and $\epsilon=0.2$. Mass scales like $M_{_d}<10^{14}$ seems not to represent growth rate today in correspondence to observations ($f_{_0}=0.5$) \cite{Linder2005}. (b) Evolution of the linear growth rate since the recombination era. The Lahav approximation from equation \eqref{LahavEquation} is also plotted. These curves are only for $M_{_d}=10^{14} M_\odot$ and consider the WMAP result set $\Omega_{_{b0}}=0.0458$, $\Omega_{_{d0}}=0.229$ and $h=0.702$ \cite{Komatsu2011_192}.}
\end{figure}

Based on the results from seven-year WMAP data set (WMAP7 - \cite{Komatsu2011_192}), we assume a spatial flat universe with density parameter values for current baryonic matter $\Omega_{_{b0}}= 0.0458$ and dark matter $\Omega_{_{d0}}=0.229$, and the Hubble constant $H_0= 70.2~km~s^{-1}~Mpc^{-1}$. As initial conditions for our numerical code the evolution of the fluctuations start when 90\% of all primordial hydrogen already recombined (when the radiation temperature $T_\gamma \simeq 4000 \, K $ and $ z \simeq 1500 $).
The inputs for the initial contrast of baryonic density  is $\delta_d^{rec} = 0$ (because we do not expect any initial fluctuation of baryonic matter at this time due to its recently tight interaction with the CMB photons) and $\delta_{_{d}}^{rec}=10^{-5}$ (after matter radiation equality, dark matter fluctuations grew little and we supposed that they are not much bigger than that of the CMB). The amount of dark mass of the fluctuation is also an initial input of the numerical calculation and the value that we used was $ 10^{14} \, M_\odot$, which mimics a cluster of galaxies and almost a linear fluctuation. Figure~\ref{FigGrowth1}a shows clearly this point: fluctuations with $M_d < 10^{14} \, M_\odot$ fail to represent $f(z=0)$ corresponding to the observations (see table~\ref{tb02} for the observational data).

The evolution of the linear growth rate since recombination is shown in figure~\ref{FigGrowth1}b. For the sake of comparison, the Lahav approximation from equation \eqref{LahavEquation} is also plotted in the same figure (dashed curve).

\begin{figure}[hptb] 
\vspace{0.5truecm}
\centering 
\hspace{0.4 truecm}\includegraphics[width=0.44\textwidth,trim=90 90 90 90]{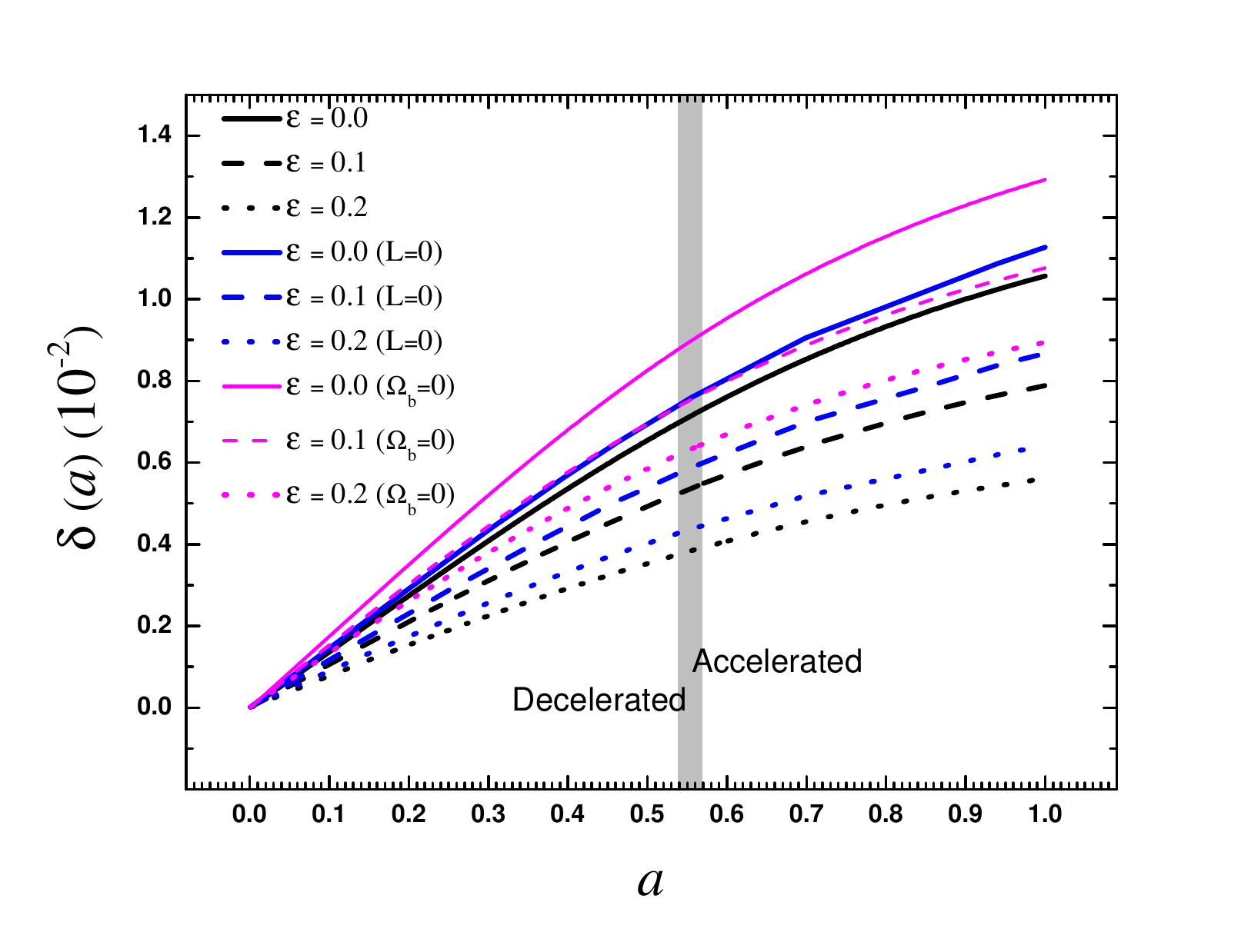}
\vspace{.5truecm}
\caption{\label{fig:1} Evolution of the density contrast of baryon and dark matter ($\delta(a)=\delta_{_b}(a)+\delta_{_d}(a)$) since the recombination until today ($a=1$) for $\epsilon = 0 $, $0.1$, $0.2$.  Black curves specifies the models which take into account dark plus baryonic matter and the physical mechanisms; blue curves specifies those with dark matter plus baryonic matter without physical mechanisms; and rose curves specifies those with only dark component. The vertical band includes the redshift of transition of the three models. For this figure it was assumed $\Omega_{_{d0}}=0.229$, $\Omega_{_{b0}}=0.0458$ and $h=0.702$.}
\end{figure}

The curve  signed $\epsilon = 0.0$ (solid curve) coincides with the $\Lambda$CDM model, and it is quite similar to Lahav approximation (long-dashed curve), at least for $0 \leq z \lesssim 100$. The cases $\epsilon = 0.1$ and $\epsilon = 0.2$ (pointed and dashed-dot curves, respectively) follow that to $\epsilon = 0.0$, but the amplitude is smaller as bigger as the value of  $\epsilon$. This trend was already expected because in these models with decaying vacuum the density of dark matter is smaller in higher redshifts (see equation \eqref{rod}). This reflects what is shown in figure~\ref{fig:1} of the evolution of the total density contrast. Note that, all the models showed in this figure has its present  growth rate value (i.e., $f(z=0)$) close to the observational value $\Omega_{_m}^{0.55}=(\Omega_{_{b0}}+\Omega_{_{d0}})^{0.55} \simeq 0.48$ (see \cite{Lightman1990_74} or \cite{Beutler}). The second and third column of table~\ref{tb01}, give the numerically calculated values for the today values of the growth rate (respectively with and without the baryonic component) for three classes of models with $\epsilon = 0.0$ ($= \Lambda$CDM), $0.1$, and $0.2$.

\begin{table}[!hptb]
\centering
\begin{tabular}{@{}|r|cc|@{}} \hline
$\epsilon\;\;$ &$f_{_0}$ &$f_{_0}^*$ \\
\hline
$0.0$         &$0.490$  &$0.442$ \\
$0.1$         &$0.486$  &$0.436$ \\
$0.2$         &$0.483$  &$0.430$ \\
\hline
\multicolumn{3}{@{}l}{\emph{$^{\emph{$*$}}$ {\footnotesize{Stands for the case $\Omega_{_b}=0$.}}}}
\end{tabular}
\caption{\label{tb01} Today values of the numerical growth rate, with and without the baryonic component, $f_{_0}$ and $f_{_0}^*$, respectively, for three values of the parameter $\epsilon$. In this analysis $\Omega_{_{d0}}=0.229$, $\Omega_{_{b0}}=0.0458$, and $h=0.702$. }
\end{table}

\begin{figure}[hptb]
\vspace{1.5truecm}
\begin{center}
\includegraphics[width=.4\textwidth,trim=90 90 90 90]{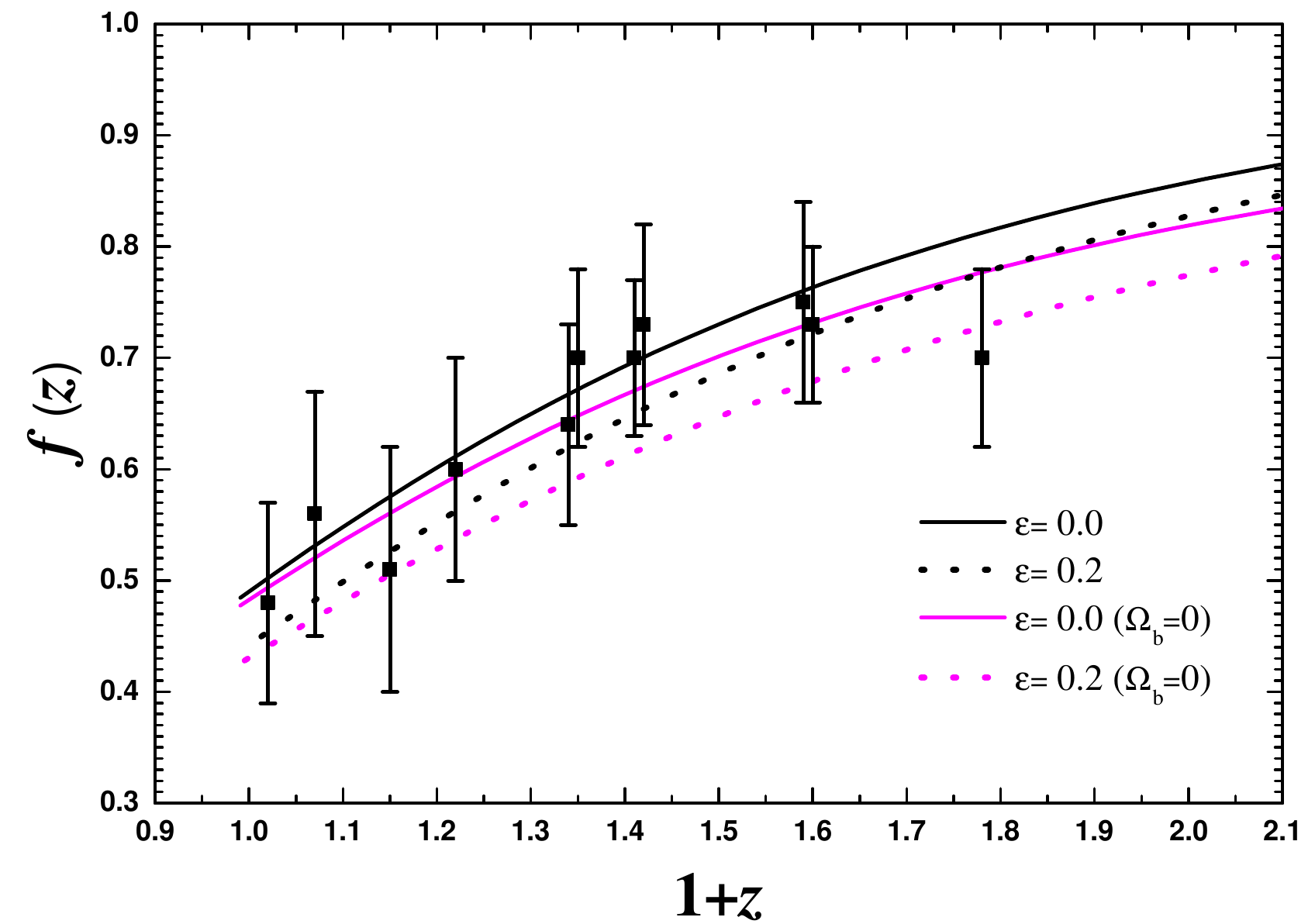}
\vspace{0.7truecm}
\caption{\label{FigGrowth2} The points with error bar are the observed data (see table~\ref{tb02}). This figure shows also the curves for $\epsilon = 0.0$  ($\equiv  \Lambda$CDM) and $\epsilon = 0.2$ with the inclusion of baryonic matter and without it. No significant difference from the curves with baryonic matter was found when the physical cooling/heating processes were absent.}
\end{center}
\end{figure}

Figure~\ref{FigGrowth2} gives the evolution of $f(z)$ in low redshifts, and shows the importance of take into account the baryonic matter.

In this figure is also superimposed the observed values of $f(z)$ and their error bars, which are tabled in table~\ref{tb02}. The theoretical curves shown that a smaller amplitude of $f(z)$ for the cases without baryonic matter. Comparing the results obtained with the observational data set, we find that these models are viable because they are within the error bars.

Figure~\ref{fig:1} shows the evolution of the total (dark plus baryonic) density contrast for the three classes of models with $\Lambda(t)$: $\epsilon = 0.0$ ($= \Lambda$CDM plus baryonic matter), $0.1$, and $0.2$. In this picture is also presented the effects of take into account dark matter and baryonic matter plus the physical mechanisms (dark curves), dark matter and baryonic matter without physical mechanisms (blue curves marked with `$L=0$') and  only the dark component (rose curves marked with `$\Omega_b=0$').
Although dark energy is not dominant during the primordial times, these curves show its influence over the growth of the perturbations. However, this result was expected because different rates of vacuum decay means different quantities of dark matter. Then, as bigger as  the $\epsilon$ parameter, less is the dark matter quantity in early times. This delays the growth of the perturbations, making the amplitude of $\delta(a)$ to be smaller for bigger $\epsilon$'s. This figure also shows the importance of the baryonic component as well the physical mechanisms taken together or alone. They delay the growth of the fluctuation. In the case with only baryonic matter without, the physical mechanisms is easy to see from equation   \eqref{normalization}, which shows that in case without baryonic matter the parameter $\tilde{\Omega}_{_{v0}}$ is bigger (because the model is flat); and when the physical mechanisms are considered the main delay source is the cooling due to the formation of molecules. For a summary of this analysis see table~\ref{tb00}.

\begin{table}[hptb]
\centering
\begin{tabular}{|l|c|c|c|c|c|c|c|}
\hline
$\epsilon$  &$\Omega_{_{d0}}$   &$\Omega_{_{b0}}$   &$\Omega_{_{v0}}$
 &Cooling    &$\delta_{_{d0}}~(10^{-2})$ &$\delta_{_{b0}}~(10^{-2})$
&$\delta_{_0}~(10^{-2})$ \\
\hline \hline
$0.0$       &$0.229$            &$0.0458$           &$0.7252$
 &yes        &$0.53$                     &$0.53$     &$1.06$ \\
$0.0$       &$0.229$            &$0.0458$           &$0.7252$
 &no         &$0.56$                     &$0.56$     &$1.13$\\
$0.0$       &$0.229$            &$0$                 &$0.7710$
 &--        &$0.65$                     &$0.65$      &$1.29$ \\
\hline
$0.1$       &$0.229$            &$0.0458$           &$0.7252$
 &yes        &$0.39$                     &$0.39$     &$0.79$ \\
$0.1$       &$0.229$            &$0.0458$           &$0.7252$
 &no         &$0.43$                     &$0.43$     &$0.87$\\
$0.1$       &$0.229$            &$0$                 &$0.7710$
 &--         &$0.54$                     &$0.54$     &$1.08$\\
\hline
$0.2$       &$0.229$            &$0.0458$           &$0.7252$
 &yes        &$0.28$                     &$0.28$     &$0.56$\\
$0.2$       &$0.229$            &$0.0458$           &$0.7252$
 &no         &$0.32$                     &$0.32$     &$0.64$\\
$0.2$       &$0.229$            &$0$                 &$0.7710$
 &--         &$0.45$                     &$0.45$     &$0.89$ \\
\hline
\end{tabular}
\caption{\label{tb00} Summary of the analysis about the figure~\ref{fig:1} discussed in the text. The final density contrast (in $a = 1$) are $\delta_{_{d0}}$, $\delta_{_{b0}}$ and  $\delta_{_0}= \delta_{_{d0}} + \delta_{_{b0}}$ of the dark matter, baryonic matter and the total density contrast of the cloud.}
\end{table}

The confidence levels result for the $\chi^2$ minimization, from equation~\eqref{chi2gf}, are showed in figure~\ref{Cosm_Tests} as black dashed curves. To delimit the degeneration of the space of parameters, the model (parameters $\Omega_{_{d0}}-\epsilon$) was also compared with two more sets of observational data. One of them is the CMB-BAO ratio and the other is the cluster gas mass fraction from X-ray data, discussed in the following section.

\section{Complementary constraints}~\label{Constraints}

\subsection{CMB-BAO ratio}~

As it is largely known, the CMB constrains from the so-called shift parameter $ \cal{R}$ \cite{elgaroy} and the BAO measurement $A(z_{BAO})$ \cite{Eisenstein2005_633} have been commonly used to constrain non-standard models, but this approach is not completely correct since these quantities are derived by using parameters close to standard $\Lambda$CDM (see \cite{Sollerman2009_703}, \cite{elgaroy}, \cite{doran}).

In order to avoid some bias in our results we follow reference~\cite{Sollerman2009_703} and use the ratio CMB/BAO, i.e., $ d_A(z_*) / D_V(z_{BAO})$ where $z_*$ is the sound horizon at the last scattering. In fact, this is a product that cancels out some of the dependence on the sound horizon size at last scattering and a more model-independent constraint can be achieved in statistical analysis. Reference~\cite{Sollerman2009_703} gives this relation for the redshifts 0.2 and 0.35 (that come from the BAO in these redshifts) which were joined four more BAO data from 6dF Galaxy Survey Redshift (6dFGS) \cite{Beutler2012_423} and in the WiggleZ Dark Energy Survey \cite{Blake2011}. These relations are presented in table~\ref{tb_01a}. The $\chi^2$ function to be minimized is
\begin{equation}
\chi_{_{CMB-BAO}}^2 = \displaystyle\sum_{i=1}^{6}\left(\frac{\left[\frac{d_{_{A}}(z_{_*})}{D_{_{V}}(z_{_{BAO,i}})}\right]_{obs}-\left[\frac{d_{_{A}}(z_{_*},\theta_{_i})}{D_{_{V}}
(z_{_{BAO,i}},\theta_{_i})}\right]_{th}}{\sigma_{_{CMB-BAOi}}}\right)^2,
\end{equation}
where the theoretical ratio CMB-BAO is

\begin{equation}\label{chi2bao}
\frac{d_{_A}(z_{_*})}{D_{_V}(z_{_{BAO}})}=\displaystyle\frac{\displaystyle\int^{z_{_*}}_{_0}\frac{dz_{_*}'}{E(z_{_*}')}}{(1+z_{_*})}
\left[\frac{E(z_{_{BAO}})}{z_{_{BAO}}\left(\displaystyle\int_{_0}^{z_{_{BAO}}}\frac{dz_{_{BAO}}'}{E(z_{_{BAO}})}\right)^2}\right]^{1/3}\;,
\end{equation}
\\
$z_{_*} = 1090 $ is the redshift assumed for the recombination, $E(z)= H(z)/H_0$ is given by equation \eqref{H}, and $z_{_{BAO}}$ is the redshift of the BAO peaks (table~\ref{tb_01a}).

\begin{table}[hptb]
\centering
\begin{tabular}{|c| l| r|} \hline
Sample &$z_{_{BAO}}$ &$d_{_A}(z_{_*})/D_{_V}$ \\
\hline
6dFGS      &$0.106$     &$30.95 \pm 1.50$\\
2dFGRS      &$0.2$      &$17.55 \pm 0.65$\\
SDSS      &$0.35$    &$10.10 \pm 0.38$\\
WiggleZ      &$0.44$       &$8.43 \pm 0.67$\\
WiggleZ      &$0.6$     &$6.69 \pm 0.33$\\
WiggleZ      &$0.73$      &$5.45 \pm 0.31$ \\
 \hline
 \end{tabular}
\caption{\label{tb_01a} The observational CMB-BAO ratio data set.}
\end{table}

The confidence levels are showed in figure~\ref{Cosm_Tests} as dotted red curves.

~

\subsection{Gas mass fraction constraint}~

For the other complementary observational comparison, we consider the data set of the baryonic content of 52 X-ray luminous galaxy clusters observed with Chandra in the redshift range $0.3<z<1.273$ provided by Ettori et al. \cite{Ettori2009_501}.

Using equations (8) and (11) from \cite{Ettori2009_501}, we have

\begin{equation}\label{fgasmodelo}
f_{gas}^{model} = \dfrac{B\Omega_{_b}}{[1.18-0.012kT_{gas}]\Omega_{_M}}
\end{equation}
\\
where the depletion parameter $B=0.874\pm0.0023$, the baryonic density parameter $\Omega_{b0} = 0.0458\pm0.0016$ and $kT_{gas}$ is the gas temperature of each cluster .

However, the X-ray gas mass fraction values were determined for $\Lambda$CDM ($ H_0 , \Omega_M , \Omega_\Lambda $) = ($70 \, km\, s^{-1}\, Mpc^{-1}$), but since $f_{gas} \propto d_A^{3/2}$ (see \cite{sasaki}), the model function is defined by

\begin{equation}
f_{gas}^{model} = \dfrac{B\Omega_{b}}{[1.18-0.012kT_{gas}]\Omega_{_M}}\left(\dfrac{d_{A}^{\Lambda CDM}}{d_{A}^{model}}\right)^{3/2}\;.
\end{equation}
\\
The distance ratio $(d_{A}^{\Lambda CDM}/d_{A}^{model})$ accounts for deviations in the geometry of the universe from the $\Lambda$CDM model, where

\begin{equation}\label{dalcdm}
d_{A}^{\Lambda CDM} = \dfrac{c}{(1+z)H_{70}}\int_{0}^{z}{\dfrac{dz}{\sqrt{0.3(1+z)^{3}+0.7}}}
\end{equation}and

\begin{equation}\label{modelo}
d_{A}^{model} = \dfrac{c}{(1+z)H_{0}}\int_{0}^{z}{\dfrac{dz}{H(z)}}
\end{equation}
\\
where $H(z)$ is the Hubble parameter of the model given by equation \eqref{H}.

We find the constraints on the model parameters by minimizing the chi-squared function

\begin{equation}\label{chi2gas}
\chi^{2}_{f_{gas}} = \sum_{i=1}^{52}{\dfrac{[f_{gas}^{model}(z_{i},\theta_{i})- f_{gas}^{obs}(z_{i},\theta_{i})]^{2}}{\sigma^{2}(i)_{f_{gas}^{model}}+\sigma^{2}(i)_{f_{gas}^{obs}}}}
\end{equation}
\\
where $\sigma^{2}(i)_{f_{gas}^{model}}$ are the propagate errors, expressed by

\begin{equation}
\sigma(i)_{f_{gas}^{model}} = f_{gas}^{model}\sqrt{0.07+\left(\dfrac{0.012kT(i)}{[1.18-0.012kT(i)]}\sigma(i)_{_{kT}}\right)^{2}}
\end{equation}
\\
and $\sigma^{2}(i)_{f_{gas}^{obs}}$ are the errors for the $\Lambda$CDM data.

A similar analysis using gas mass fraction was done by first reference in \cite{Gon?alves}, but they used the $f_{gas}$ sample calculated in reference \cite{Allen}.

\begin{figure}[!htbp]
\vspace{1.truecm}
\centering 
\hspace{-1. truecm}\includegraphics[width=.45\textwidth,trim=90 90 90 90]{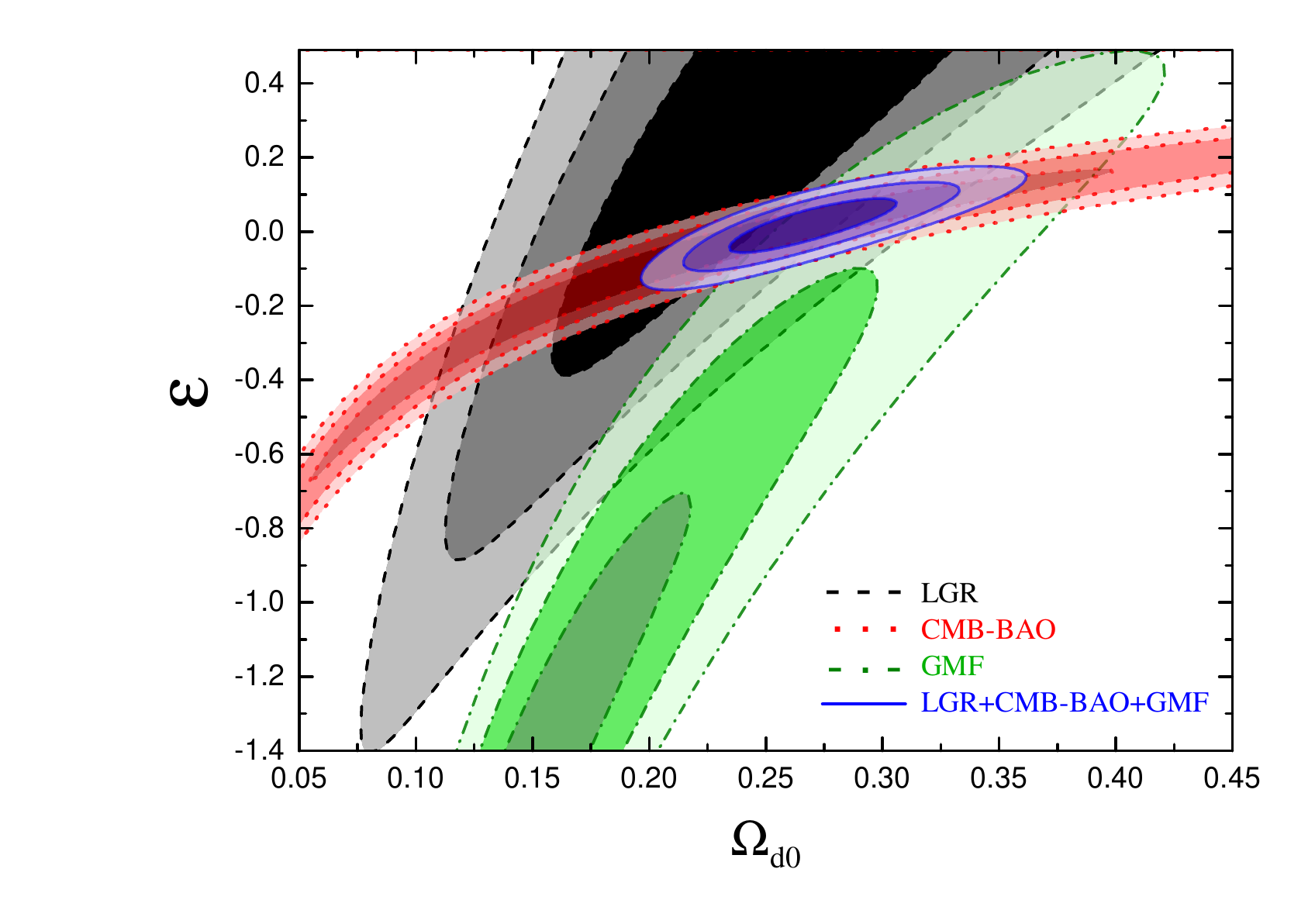}
\vspace{.6truecm}
\caption{\label{Cosm_Tests} Superposition of the confidence level contours of $68.3\%$, $95.4\%$ and $99.73\%$ in the $\Omega_{_{d0}}-\epsilon$ plane from the linear growth rate (LGR), CMB-BAO ratio set (CMB-BAO) and the gas mass fraction (GMF), as well as their combination. The best fits and error bars are to $1\sigma~(\Delta\chi^2=1)$ shown in table~\ref{tb_05}.}
\end{figure}

The confidence levels are showed in figure~\ref{Cosm_Tests} as green dashed-dot curves. The joint analysis is given by
\begin{equation}\label{chi2joint}
\chi_{_{joint}}^2 = \chi_{_{gf}}^2 + \chi_{_{CMB-BAO}}^2 + \chi^{2}_{fgas} \;\; .
\end{equation}
This analysis is showed in figure~\ref{Cosm_Tests} as continuous blue curves.

Figure~\ref{Cosm_Tests} shows that the regions representing the constraints from growth rate and gas mass fraction measurements on the parameter space $\Omega_{_{d0}}-\epsilon$ are almost parallel, while the CMB-BAO measurements are approximately orthogonal to the last two measurements constraints.
We also note that, notwithstanding the growth rate and gas mass fraction measurements come from different data sets, the first represents approximately the density contrast and the second the gas mass fraction in the cluster. This can mean that these data are measuring almost the same thing, which could explain why the confidence levels are parallel.

The best fit values for each data set and for the joint analysis are presented in Table~\ref{tb_05}.
\begin{table}[!htbp]
\centering
\begin{tabular}{|c|c|c|c|} \hline
Test                        &$\Omega_{_{d0}}$   &$\epsilon$     &$\chi_{_{min}}^2/\nu$ \\
\hline
LGR   &$0.254^{+0.054}_{-0.066}$      &$~0.35^{+0.13}_{-0.49}$        &$0.454$\\
CMB-BAO         &$0.193^{+0.125}_{-0.097}$            &$-0.11^{+0.19}_{-0.30}$  &$0.815$\\
GMF             &$0.137^{+0.048}_{-0.036}$            &$-1.68^{+0.64}_{-0.68}$  &0.986\\
LGR+GMF+CMB-BAO   &$0.269^{+0.023}_{-0.023}$          &$~0.02^{+0.04}_{-0.05}$  &0.999\\
 \hline
 \end{tabular}
\caption{\label{tb_05} Limits to $\Omega_{_{d0}}$ and $\epsilon$ with error bars standing for $1\sigma$. LGR, CMB-BAO and GMF in the first column refer to Linear Growth Rate, Cosmic Microwave Background radiation - Baryon Acoustic Oscillations ratio and Gas Mass Fraction, respectively.}
\end{table}

The results that we found are similar to that presented in references \cite{Alcaniz2005_72} and \cite{Gon?alves}. This mean that these results reinforces the conclusion already found by other authors about the possibility of a small, but measurable, deviation from the standard $\Lambda$ dynamics (formally equivalent to the case $\epsilon = 0$).
%
%
\section{Conclusions}~ \label{Conc}

In this paper we have studied the observational consequences of the decaying vacuum scenario proposed by Wang and Meng \cite{Wang2005_22} considering the baryonic component, whose importance was pointed by Alcaniz and Lima \cite{Alcaniz2005_72}, as well we have done one more slightly modification in the original model by considering the interaction between dark and baryonic matter components.

In order to enhance the importance of the baryonic component, two of the three analysis done was based in the observational data related with the baryonic mass, namely the `growth rate' and `gas fraction in clusters of galaxies'. The third analysis was based in observational CMB-BAO rate, which proved to be a very useful complementary analysis because its confidence levels are transverse to the two first.

The theoretical counterpart of the growth rate was new and much more sophisticated than those we found in the literature. In this calculation, the baryonic and dark matter was considered as two fluids coupled and the main physical mechanisms present during and after recombination was also taken into account.

In conclusion, our statistical analysis reinforces the conclusion already found by others authors about the possibility of a small but measurable deviation from the standard $\Lambda$CDM.

%
%

\begin{acknowledgments}
The author H. T. C. M. Souza is very grateful for the Brazilian
agency CAPES. Also H. T. C. M. Souza is very grateful for the PDJ fellowship of the INCT INEspa\c{c}o/CNPq
for financial support. The author E. P. Bento is very grateful for the PNPD fellowship of the CAPES
for financial support.
\end{acknowledgments}
%
%
\appendix
\section{Temperature of the baryonic matter cloud} \label{Ap_A}~

From recombination until the beginning of the collapse of the first structures, the baryonic matter (protons, electrons and
hydrogen) can be treated as an ideal gas whose equation of state are given by
\begin{equation}
P=N_{_A}k_{_B}\rho_{_{cb}}(1+x_{_e})T_{m}\;\; , \label{P}
\end{equation}
where $T_{_m}$ is the temperature of the baryonic matter, $k_{_B}$ is the Boltzmann's constant, $N_{_A}$ the Avogadro's number, $\rho_{_{cb}}$ is the density of baryonic matter and the parameter $x_{_e}$ is the degree of ionization of the matter (in the beginning of the recombination $x_{_e}=1$). To simplify, in this work we have used the recombination given by \cite{Peebles1968_153}, modified to include the collisional ionization among electrons and hydrogen atoms \cite{Defouw1970_161}.

The energy equation is written as
\begin{equation}
\frac{dU}{dt}=-L+\frac{P}{\rho^2_{_{cb}}}\frac{d \rho_{_{cb}}}{dt}\;\;, \label{dUdT}
\end{equation}
where $U$ is the internal energy of the gas which is (per unit of mass $m$),
\begin{equation}
U=\frac{3}{2}N_{_A}k_{_B}T_{_m}(1+x_{_e})\;\; , \label{U}
\end{equation}
and $L$ is the cooling function which take into account the physical cooling/heating processes acting on the baryonic matter, and it is given by
\begin{equation}
 L=-k_{_B}N_{_A}T_{_m}\frac{dx_{_e}}{dt}-\displaystyle{\frac{4k_{_B}\sigma_{_T}bT_{_\gamma}^{4}x_{_e}}{m_{_p}m_{_{e}}c}(T_{_m}-T_{_\gamma})}+L_{H_{2}}+
 L_{\alpha}\;\; ,  \label{L}
\end{equation}
which includes recombination, photoionization and collisional ionization (first term), photon cooling (heating, second term), molecular hydrogen cooling and Lyman-$\alpha$ cooling (third and forth terms, respectively). In this equation, $m_{_p}$ and $m_{_e}$ are the proton and electron masses, respectively; the speed of light is $c$, $\sigma_{_T}$ the Thomson cross section and $b=4\sigma/c$, where $\sigma$ is the Stefan-Boltzmann's constant. For more details about these terms see ref~\cite{sandra}.

Using equations \eqref{P} and \eqref{U} into \eqref{dUdT}, and  $\rho_{_{cb}} =\rho_{_{b0}}(1+\delta_{_b})$, we obtain the equation that governs the evolution of the temperature of the matter inside the cloud
\begin{equation}
\frac{dT_{_m}}{dt}=-T_{_m}\left[\frac{\dot x_{_e}}{1+x_{_e}}+2\frac{\dot a}{a}-\frac{2}{3}\frac{\dot \delta_{_b}}{1+\delta_{_b}}\right]-\frac{2}{3}\frac{L}{N_{_A}k_{_B}(1+x_{_e})}\;\;.\label{Tmb}
\end{equation}

The evolution of baryonic matter temperature of the Universe is the same equation \eqref{Tmb} disregarding the term with $\delta_{_b}$ (the last term in brackets).

For more details, the basic equations for the physical mechanisms were presented in various papers of de Araujo \& Opher  \cite{Opher1988_231}.

%

\end{document}